\documentclass[10pt]{article}
\usepackage[a4paper,margin=15mm,heightrounded,includefoot]{geometry}

\newcommand{\receiveddate}[1]{}
\newcommand{\reviseddate}[1]{}
\newcommand{\accepteddate}[1]{}
\newcommand{\publisheddate}[1]{}
\newcommand{\currentdate}[1]{}
\newcommand{\doiinfo}[1]{}

\newenvironment{IEEEkeywords}{\paragraph*{Index Terms—}}{}

\usepackage{authblk}

\usepackage{amsmath,amssymb,amsfonts}
\usepackage{graphicx}
\usepackage{cite}
\usepackage{algorithm}
\usepackage{algorithmic}
\usepackage{xcolor}
\usepackage{siunitx}
\usepackage{hyperref}
\hypersetup{hidelinks}

\usepackage{tikz}
\usetikzlibrary{arrows.meta,positioning,calc,fit,backgrounds,decorations.pathreplacing,shapes.geometric}

\title{Beyond Imaging: Vision Transformer Digital Twin Surrogates for 3D+T Biological Tissue Dynamics}
\author[1]{Kaan Berke Ugurlar}
\author[2]{Joaquín de Navascués}
\author[1]{Michael Taynnan Barros\thanks{Corresponding author: michael.barros@essex.ac.uk}}
\affil[1]{School of Computer Science and Electronic Engineering, University of Essex, UK}
\affil[2]{School of Life Sciences, University of Essex, UK}
\date{} 

\begin{document}
\maketitle

\begin{abstract}
Understanding the dynamic organization and homeostasis of living tissues requires high-resolution, time-resolved imaging coupled with methods capable of extracting interpretable, predictive insights from complex datasets. Here, we present the Vision Transformer Digital Twin Surrogate Network (VT-DTSN), a deep learning framework for predictive modeling of 3D+T imaging data from a biological tissue. By leveraging Vision Transformers pretrained with DINO (Self-Distillation with NO Labels) and employing a multi-view fusion strategy, the VT-DTSN learns to reconstruct high-fidelity, time-resolved dynamics of a \textit{Drosophila} midgut tissue while preserving morphological and feature-level integrity across imaging depths. The model is trained with a composite loss prioritizing pixel-level accuracy, perceptual structure, and feature-space alignment, ensuring biologically meaningful outputs suitable for in silico experimentation and hypothesis testing. Evaluation across layers and biological replicates demonstrates the VT-DTSN’s robustness and consistency, achieving low error rates and high structural similarity while maintaining efficient inference capability through model optimization. This work establishes VT-DTSN as a feasible, high-fidelity surrogate for cross-timepoint reconstruction, for studying tissue dynamics, enabling computational exploration of cellular behaviors and homeostasis to complement time-resolved imaging studies in biological research.
\end{abstract}


\section{INTRODUCTION}

High‑resolution 3D+T confocal imaging of the \textit{Drosophila} midgut enables quantitative study of epithelial homeostasis and regeneration. However, exploiting its depth‑varying signal and heterogeneous cellular architecture remains computationally challenging \cite{miguel2018anatomy,jiang2016intestinal,jiang2012intestinal}. The sheer complexity and volume of this data present significant analytical challenges. A typical acquisition comprises several optical sections per Z‑stack at high resolution with micrometric voxel size, repeated across independent biological replicates and time points. Depth‑dependent scattering and photobleaching reduce SNR in deeper layers, while phototoxicity limits repeated imaging of the same specimen. These constraints create partially observed, noisy 3D+T volumes where long‑range spatial dependencies (across layers) and temporal consistency must be recovered from limited sampling data. Traditional computational methods fall short in capturing the nuanced spatial-temporal patterns inherent in tissue dynamics, limiting our ability to fully leverage this wealth of information.

Standard convolutional neural networks (CNNs) pipelines emphasize local receptive fields and under‑represent long‑range cross‑layer structure \cite{huang2025deep,arledge2022deep}. recurrent neural networks (RNNs) and temporal CNNs mitigate this only partially and are costly to scale on volumetric data \cite{hinrichsen2024using}. Moreover, models trained on narrowly curated imaging conditions often fail to generalize across biological replicates with variable contrast, labeling, and depth attenuation. The specialized nature of experimental tissue imaging datasets leads to difficulties in model generalization. Furthermore, the sophisticated spatial-temporal behaviour of tissue processes demands more than what standard CNN and RNN architectures can offer. What is currently missing is a data‑driven surrogate that can reconstruct and predict midgut dynamics from sparse or noisy 3D+T stacks with fidelity to morphology and features, while remaining efficient enough for near real‑time use. 

\begin{figure}
    \centering
    \includegraphics[width=\linewidth]{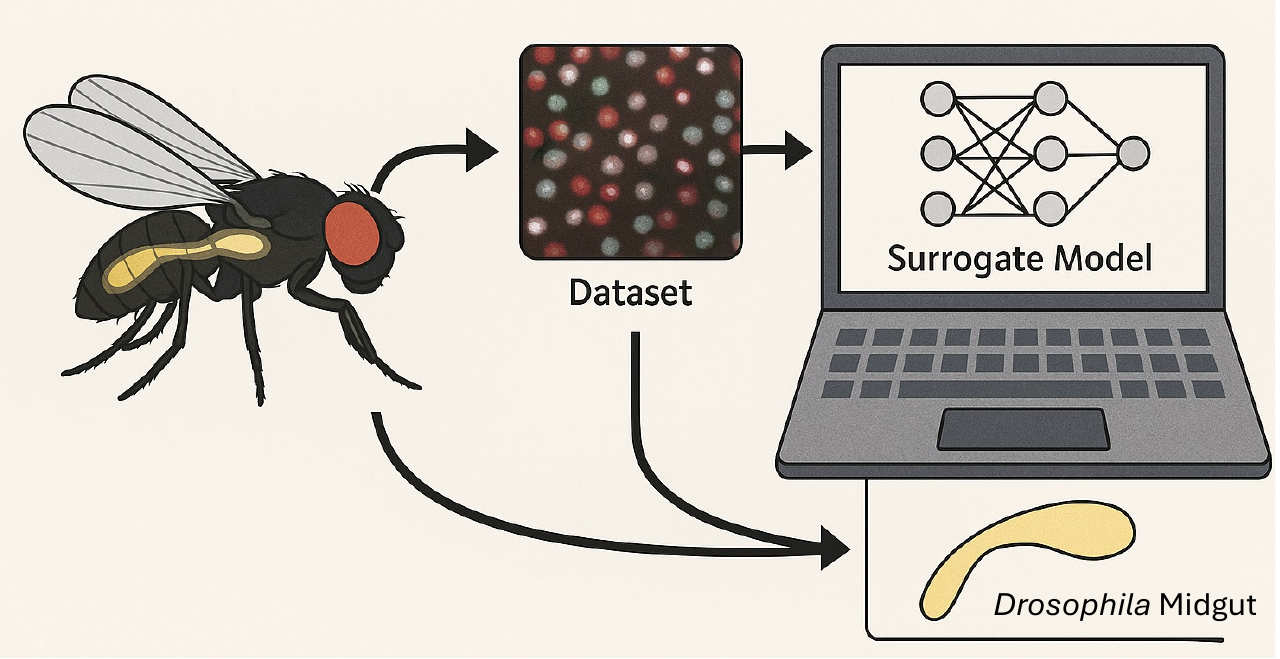}
    \caption{Conceptual schematic illustrating the Vision Transformer Digital Twin Surrogate Network (VT-DTSN) workflow for \textit{Drosophila} midgut homeostasis studies. High-resolution 3D+T imaging datasets of the midgut are acquired and input into the VT-DTSN pipeline, where Vision Transformer-based feature extraction and fusion reconstruct dynamic tissue states across time and depth.}
    \label{fig:vision}
\end{figure}

To address these limitations, we pursue a digital‑twin surrogate: a high‑fidelity, data‑driven model that reproduces observed tissue dynamics for in silico experimentation. Here, ‘\textit{digital twin}’ denotes a statistical surrogate rather than a mechanistic simulator with explicit biophysical laws. By learning from high-dimensional, time-resolved imaging data, such models can emulate the behaviour of living tissues in silico, enabling predictive simulation, hypothesis testing, and closed-loop experimental design without the constraints of physical experimentation alone. This enables integration with diverse biological datasets without requiring explicit mechanistic parameterization, making the approach adaptable to other tissues and imaging modalities. In this context, we propose the \textit{Vision Transformer Digital Twin Surrogate Network} (VT-DTSN), a framework that uses deep learning not merely for image classification or segmentation, but to generate predictive, high-fidelity reconstructions of dynamic tissue behaviours, effectively serving as a computational proxy for live biological systems. Vision Transformers, via self‑attention, model long‑range spatial relations across Z‑slices more naturally than convolutional hierarchies. 

A digital twin surrogate model of the midgut enables predictive, in silico experimentation that complements time-resolved imaging, allowing researchers to simulate tissue responses to genetic, pharmacological, or mechanical perturbations before committing to labor-intensive experimental procedures. By capturing the layered epithelial architecture and cell-type-specific dynamics across time and depth, VT-DTSN allows high-fidelity reconstruction and cross-timepoint reconstruction/prediction across replicates of midgut dynamics, offering insights into tissue organization and cellular behaviors that are impractical to measure continuously in vivo. This capability opens new pathways understanding how cellular heterogeneity and spatial structure influence gut homeostasis in \textit{Drosophila}.

Our approach harnesses the power of Vision Transformers (ViTs), utilizing their self-attention mechanisms to capture visual patterns crucial for understanding cell dynamics. We innovate further by integrating DINO (\textit{Self‑Distillation with No Labels}) pretraining, which enhances the ViTs' ability to assimilate spatial context and temporal cues crucial in cellular imaging. This methodology is complemented by a multi-view fusion strategy, augmenting the model's capability to synthesize diverse perspectives into a cohesive understanding of cellular behaviour. Self‑supervised DINO pretraining provides robust features under intensity shifts and staining variability typical of confocal imaging. A multi‑view fusion of ViT branches encourages consistency across lateral and depth cues, improving reconstructions in low‑SNR layers.

Our custom training process and loss formulation are designed to prioritize not just pixel accuracy, but also perceptual similarity and biological fidelity, which are essential for meaningful interpretations in biological research and effective surrogate modeling. Rigorous evaluation using metrics such as Mean Squared Error (MSE), Structural Similarity Index (SSIM), and Cosine Similarity ensures that the VT-DTSN aligns closely with authentic biological patterns observed in \textit{Drosophila} midgut dynamics. Furthermore, to meet the demands of real-time analysis in time-resolved imgaging experimental workflows, we implement model optimization strategies including pruning and mixed-precision inference, resulting in a computationally efficient and high-fidelity surrogate model. Recent progress in self‑supervised ViT pretraining and mixed‑precision inference makes it feasible to train robust volumetric surrogates and to deploy them at interactive speeds in imaging workflows.

The key contributions of this work are:

\begin{itemize}
\item \textit{Development of the VT-DTSN, a ViT-based surrogate digital twin model tailored for predictive reconstruction of dynamic tissue imaging data.} This provides a computational proxy that can emulate midgut tissue behavior from limited imaging data, reducing the need for exhaustive physical experiments.
\item \textit{Implementation of DINO-based ViT pretraining to enhance feature representation aligned with the complexities of biological imagery.} This improves robustness to depth-dependent signal loss and variability in labeling intensity, allowing the model to generalize across biological replicates.
\item \textit{Introduction of a multi-view ViT fusion strategy to enrich spatiotemporal feature integration and predictive accuracy.} This ensures that morphological features are consistently reconstructed across Z-stack layers, even in low-SNR regions, preserving biologically relevant structures.
\item \textit{A custom loss formulation emphasizing biological fidelity, perceptual structure, and pixel-wise precision.} This alignment between training objectives and evaluation criteria ensures reconstructions are both numerically accurate and interpretable for downstream biological analysis.
\item \textit{Optimization of the VT-DTSN for real-time predictive analysis within experimental pipelines.} This enables near-instantaneous feedback during live imaging sessions, supporting closed-loop experimental designs and rapid hypothesis testing.
\item \textit{Comprehensive validation demonstrating the alignment of the VT-DTSN outputs with experimentally acquired biological data, enabling its use as an interpretable, high-fidelity surrogate for in silico experimentation.} This confirms the surrogate’s suitability for simulating perturbations computationally, helping prioritize which experimental conditions to pursue in vivo.
\end{itemize}

\section{LITERATURE REVIEW}

\begin{figure*}[!]

\centering
\includegraphics[width=0.8\textwidth]{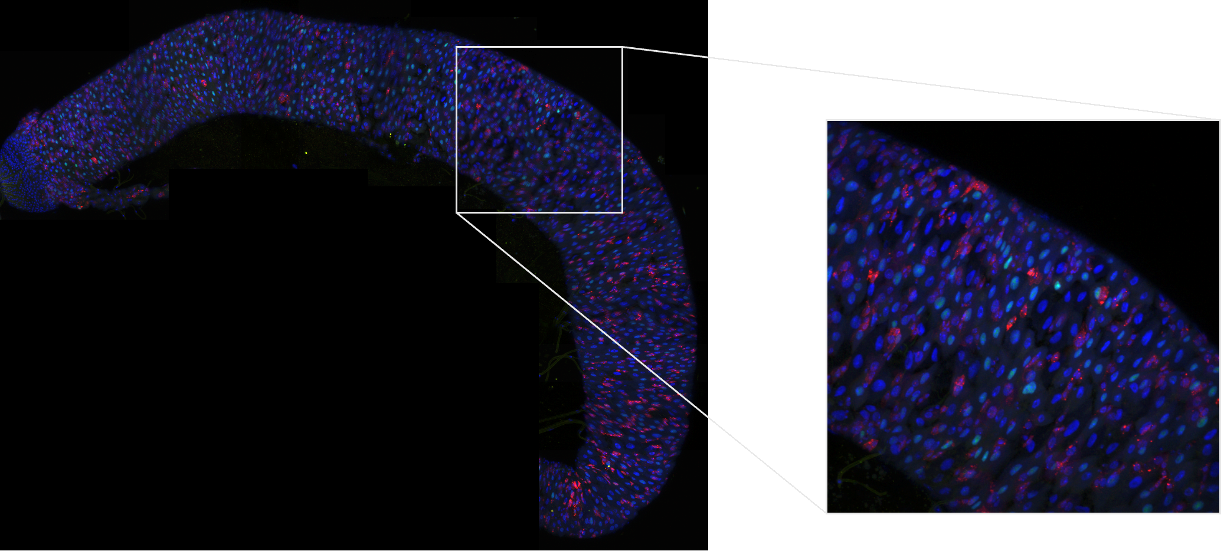}
\caption{Zoomed view of the midgut region of interest.}

\label{fig:midgut}
\end{figure*}

\begin{figure*}[!]
\centering
\includegraphics[width=0.8\textwidth]{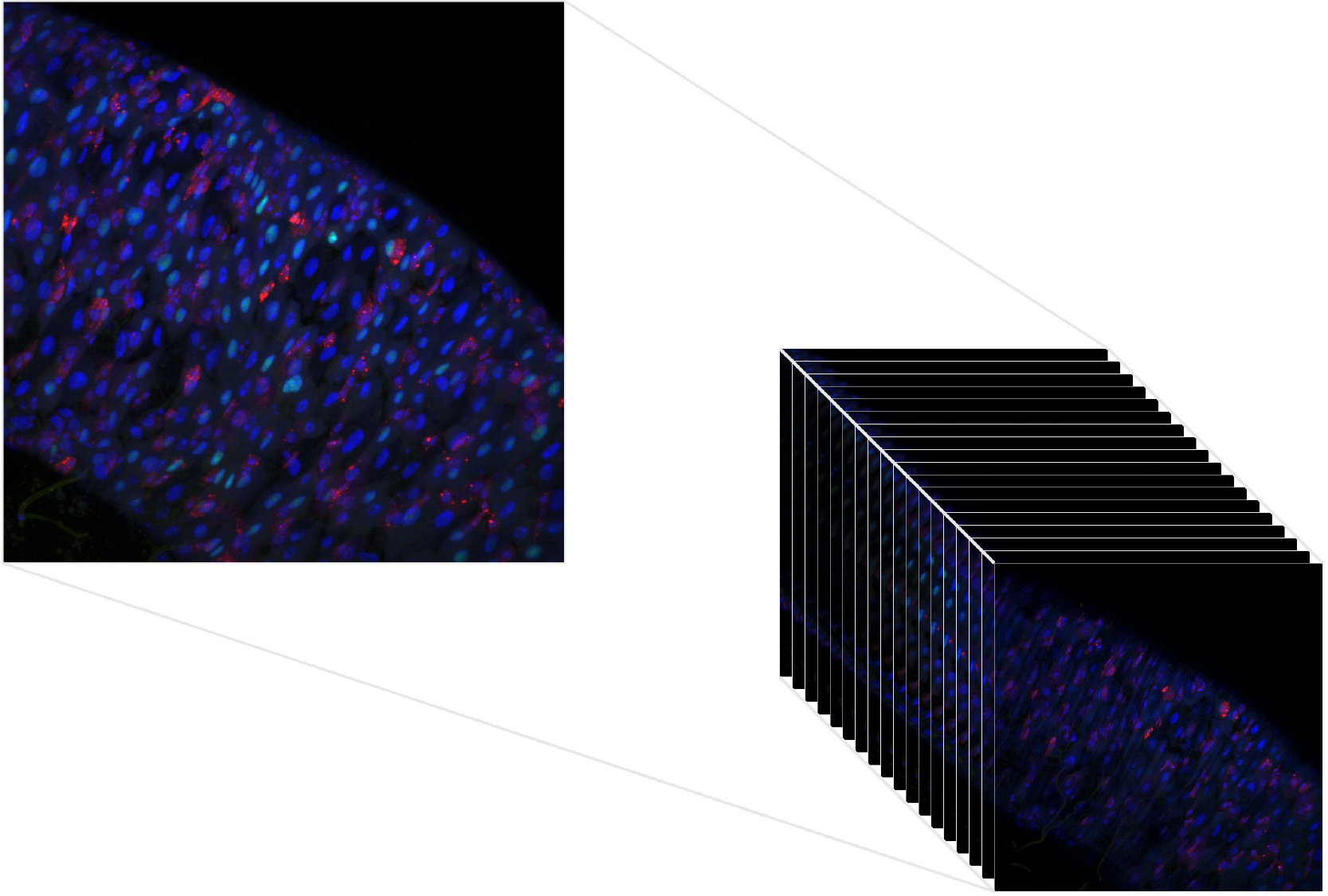}
\caption{The 18 Z-stack images creating the 3D representation.}
\label{fig:zstacks}
\end{figure*}

Traditional approaches for studying tissue structure and cellular behavior began with two-dimensional imaging of fixed samples, which enabled the characterization of cellular morphology and marker distribution but could not capture the dynamic processes of living tissues \cite{movahhedi2023predicting}. Time-resolved imaging introduced temporal resolution, allowing the observation of cellular movements and proliferation over time, yet remained largely limited to two-dimensional contexts. The transition to three-dimensional imaging using confocal and light-sheet microscopy expanded the ability to study tissue structures in greater detail, but analyzing these datasets manually or with conventional computational methods often proved infeasible due to data volume and complexity. 

The introduction of deep learning has significantly advanced the analysis of biological images, with CNNs achieving state-of-the-art performance in image segmentation, classification, and feature extraction. These models demonstrated the capacity to learn complex spatial hierarchies and enabled automated quantification in large datasets. However, the application of deep learning to dynamic 3D+T biological data remains challenging. CNNs inherently capture local spatial features and struggle with long-range dependencies \cite{huang2025deep,arledge2022deep}, while RNNs for temporal modeling are limited by issues such as vanishing gradients and high computational cost when scaling to high-dimensional data \cite{hinrichsen2024using}. Physics-Informed Neural Networks are a class of deep learning models that integrate physical laws, into the training process, with great results in allowing the construction of surrogate models that honor physical constraints and can work with limited data; successfully applied in biomedical contexts like blood flow and biomechanics \cite{movahhedi2023predicting}. PINNs also face challenges (discussed later) in scaling to very large problems or handling complex multi-physics interactions \cite{movahhedi2023predicting}
Addressing these challenges requires models that can handle complex spatiotemporal dependencies while remaining interpretable and generalizable across diverse biological contexts.

Recent advances in Vision Transformers (ViTs) have introduced a promising alternative to traditional CNN-, RNN-based and PINN-based models for biological image analysis \cite{wang2024crnn}. ViTs utilize self-attention mechanisms to capture long-range spatial dependencies, making them well-suited for high-dimensional biological imaging where context across larger spatial scales is essential. Emerging studies have demonstrated the potential of ViTs in segmentation and feature extraction tasks in biomedical imaging \cite{wang2024crnn}; however, their application in modeling dynamic, time-resolved tissue data remains limited. Integrating ViTs with strategies such as DINO pretraining and multi-view data fusion offers the potential to enhance feature learning and generalization in biological contexts while maintaining interpretability \cite{kim2025self}. This progression from static imaging analysis to dynamic, predictive modeling using ViTs represents a significant step toward creating computational tools capable of functioning as surrogate models of biological systems.

In this work, we build upon these advances by employing Vision Transformers within the Vision Transformer Digital Twin Surrogate Network (VT-DTSN), leveraging DINO pretraining and a multi-view fusion strategy to enable high-fidelity, predictive reconstruction of 3D+T imaging data from extracted \textit{Drosophila} midgut tissue. We address the current limitations in dynamic tissue analysis and supports the development of in silico experimental platforms for studying cellular dynamics and tissue homeostasis with high spatial and temporal resolution.

\section{METHODS}

\subsection{Digital Twin Surrogate Model}

To create surrogate models is to replace or accelerate traditional physics-based simulations (or complicated analytical models) with a neural network that produces similar outputs in a fraction of the time. This is attractive in medicine and biology because it can enable real-time analysis or rapid what-if simulations that were previously infeasible during clinical decision-making. Constructing a surrogate model with deep learning generally involves the following: first, generate or collect a dataset of input–output examples from the process to be emulated (this might be simulation data from many runs of a finite element model, or images paired with known parameter maps, etc.). Then, choose a suitable network architecture and train it to learn the mapping from inputs to outputs. Once trained, the neural network serves as a reduced-order model of the original process – it can instantly produce results given new inputs, whereas the traditional model might take minutes or hours.

In our setting, constructing the surrogate proceeds as follows: we assemble paired examples from 3D+T confocal imaging of the \textit{Drosophila} midgut—inputs are multi-view z-stack frames at time \(t\) (and optionally \(t\!-\!1\)) and outputs are the corresponding high-fidelity target stack at \(t\) or the cross-timepoint reconstruction/prediction across replicates at \(t+\Delta t\).
We then fine-tune a multi-branch Vision Transformer (three DINO-pretrained ViT encoders for left/mid/right views) with a lightweight fusion--reconstruction head to learn this mapping, optimizing a composite loss (MSE + SSIM + cosine similarity) to balance pixel accuracy, structural preservation, and feature alignment.
After training—and with pruning/INT8 quantization for deployment—the VT-DTSN acts as a reduced-order \emph{digital-twin surrogate} of the imaging process: given new, potentially sparse or noisy stacks, it rapidly produces depth- and time-consistent reconstructions/predictions.

\subsection{Data Collection and Preprocessing}

\textit{Drosophila} midgut samples were extracted following established dissection protocols, isolating intact midgut tissue while preserving its luminal architecture and cellular viability. Following GFP induction, midguts were dissected and imaged immediately from separate flies at each time point (days 4, 8, and 12 post-induction). Thus, time points correspond to different biological specimens and do not involve maintaining individual guts ex vivo for repeated imaging. Imaging was performed using a Zeiss LSM confocal microscope equipped with a 40× oil immersion objective (NA 1.2) at a spatial resolution of 512×512 pixels with a voxel size of 0.625x0.625x1$\mu$m. Z-stacks comprising 18 optical sections spanning the epithelial depth were acquired for each sample on days 4, 8, and 12 post-extraction, generating high-resolution time-resolved 3D datasets across eight biological replicates, each representing an independent fly midgut extraction. Each timepoint corresponds to an independent specimen. There is no repeated imaging of the same midgut. ‘Temporal’ therefore denotes cross-sectional dynamics across biological replicates rather than within-specimen time-lapse.

To ensure data quality and consistency prior to training, we applied a systematic preprocessing pipeline. First, raw fluorescence images were denoised using a combination of median filtering (kernel size 3) to suppress salt-and-pepper noise and Gaussian filtering ($\sigma$=1.0) to reduce high-frequency fluctuations while preserving edge structures critical for cell boundary and tissue architecture interpretation. Additional optional tests with anisotropic diffusion filtering were conducted but ultimately excluded to prevent oversmoothing of fine morphological features. Pixel intensities were normalized using min-max normalization, scaling each image to the [0,1] range while preserving the original distribution and retaining high-intensity outlier pixels corresponding to marker-positive cellular structures. This approach ensured consistent dynamic range alignment across all samples and timepoints without compromising biologically meaningful signal variability.

Data were organized and split into training, validation, and test sets using a 70/15/15 ratio, ensuring that entire biological replicates were allocated to a single split to prevent data leakage and to evaluate generalization across distinct samples. Each split maintained a consistent distribution of imaging timepoints and Z-stack layers, ensuring that the training dataset captured a representative diversity of midgut morphologies while the validation and test sets provided robust, independent evaluation of the model’s predictive performance across varying spatial and temporal contexts. This curated dataset, spanning over 432 Z-stacks across eight midgut extractions, provided a comprehensive and reproducible foundation for training and evaluating the Vision Transformer Digital Twin Surrogate Network in reconstructing dynamic tissue behavior in silico.

\subsection{Neural Network Architecture}

\begin{figure*}[t!]
\centering
\begin{tikzpicture}[
    font=\large,
    box/.style={draw, rounded corners, minimum width=3.6cm, minimum height=1.2cm, align=center},
    small/.style={draw, rounded corners, minimum width=2.6cm, minimum height=0.9cm, align=center},
    tinyb/.style={draw, rounded corners, minimum width=2.2cm, minimum height=0.75cm, align=center},
    input/.style={box, fill=yellow!10},
    pre/.style={small, fill=gray!10},
    view/.style={small, fill=teal!10},
    vit/.style={box, fill=blue!10},
    fusion/.style={box, fill=orange!15},
    recon/.style={box, fill=orange!10},
    concat/.style={draw, diamond, minimum width=2.2cm, minimum height=1.1cm, fill=green!10, align=center},
    output/.style={draw, ellipse, minimum width=3.8cm, minimum height=1.2cm, fill=red!10, align=center},
    loss/.style={tinyb, fill=purple!10},
    note/.style={align=left},
    arr/.style={-{Latex[length=2mm]}, thick},
    darr/.style={-{Latex[length=1.6mm]}, dashed, thick},
    node distance=1.4cm and 1.8cm,
    scale=0.78, every node/.style={transform shape}
]

\node (input) [input] {Z-stack (e.g., 18$\times$512$\times$512)};
\node (prep)   [pre, below=of input] {Denoise (median3 + Gauss $\sigma{=}1$)\\Normalize to [0,1]};
\node (viewgen) [small, fill=teal!5, below=of prep, align=left] {Multi-view construction:\\
\quad View\_L: e.g., left crop / $(z\!-\!1{:}z)$\\
\quad View\_M: center / $(z)$\\
\quad View\_R: right crop / $(z{:}z\!+\!1)$};

\node (vitLeft) [vit, below left=1.3cm and 2.9cm of viewgen] {ViT\_left\\\small ViT-B/DINO\\\small $P{=}8,\ D{=}768,\ L{=}12,\ H{=}12$};
\node (vitMid)  [vit, below=1.3cm of viewgen]               {ViT\_mid\\\small ViT-B/DINO\\\small $P{=}8,\ D{=}768,\ L{=}12,\ H{=}12$};
\node (vitRight)[vit, below right=1.3cm and 2.9cm of viewgen]{ViT\_right\\\small ViT-B/DINO\\\small $P{=}8,\ D{=}768,\ L{=}12,\ H{=}12$};

\node[below=0.25cm of vitMid, note] (vitnote) {\small Patchify $\rightarrow$ embed $\rightarrow$ SA blocks;\\[-1pt]\small classifier head removed; features pooled from last block.};

\draw[arr] (input) -- (prep);
\draw[arr] (prep) -- (viewgen);
\draw[arr] (viewgen.west) -| (vitLeft.north);
\draw[arr] (viewgen) -- (vitMid.north);
\draw[arr] (viewgen.east) -| (vitRight.north);

\node (concat) [concat, below=2.0cm of vitMid] {Concat\\\scriptsize $[3{\times}D]\ \Rightarrow\ 2304$};
\node (fusion) [fusion, below=of concat, align=center] {Fusion head (MLP)\\
\small Linear($3D{\rightarrow}D_f$)–ReLU–Linear($D_f{\rightarrow}D$)\\
\small e.g., $D_f{=}1024$};
\node (recon)  [recon, below=of fusion, align=center] {Reconstruction head\\
\small Upsampler/Decoder (Conv/PixelShuffle)\\
\small $\rightarrow$ predicted slice/stack};

\draw[arr] (vitLeft) -- (concat);
\draw[arr] (vitMid) -- (concat);
\draw[arr] (vitRight) -- (concat);
\draw[arr] (concat) -- (fusion);
\draw[arr] (fusion) -- (recon);

\node (output) [output, below=of recon] {Predicted reconstruction};
\node (loss1) [loss, right=3.6cm of output] {$\mathcal{L}_{\mathrm{MSE}}$\\\small weight $\alpha$};
\node (loss2) [loss, above=0.9cm of loss1] {$\mathcal{L}_{\mathrm{SSIM}}$\\\small weight $\beta$};
\node (loss3) [loss, below=0.9cm of loss1] {$\mathcal{L}_{\mathrm{Cosine}}$\\\small weight $\gamma$};

\draw[arr] (recon) -- (output);
\draw[arr] (output) -- (loss1);
\draw[arr] (output) -- (loss2);
\draw[arr] (output) -- (loss3);

\node (inset) [draw, dashed, rounded corners, fill=gray!10, minimum width=5.3cm, minimum height=3.2cm, above right=4.0cm and 3.9cm of vitMid, align=center] {};
\node at (inset.north) [above, font=\large\bfseries] {ViT token processing};
\node[align=left] at ($(inset.center)+(0,0.4)$) {\small Patch $P{=}8{\times}8$ $\Rightarrow$ tokens $N$\\
\small + positional enc. (absolute)\\
\small $L{=}12$ encoder blocks, $H{=}12$ heads};
\node[align=left] at ($(inset.center)+(-0.05,-0.55)$) {\small Feature readout: mean pool last block\\
\small (no classifier head)};

\draw[darr] (inset.east) -- ++(+0.6,0) |- (vitRight.east);

\end{tikzpicture}
\caption{VT-DTSN architecture with view construction, ViT hyperparameters, fusion/reconstruction heads, and composite loss. Three views derived from each input stack feed DINO-pretrained ViT branches (patch size $P$, embedding dimension $D$, layers $L$, heads $H$). Branch features are concatenated and fused via an MLP before reconstruction to a predicted slice/stack. The composite loss ($\alpha\,\mathcal{L}_{\mathrm{MSE}}{+}\beta\,\mathcal{L}_{\mathrm{SSIM}}{+}\gamma\,\mathcal{L}_{\mathrm{Cosine}}$) aligns pixel accuracy, structural similarity, and feature-space consistency. Configure multi-view to match your data: \emph{crops} (left/center/right), \emph{adjacent Z-slices} (e.g., $[z\!-\!1,z]$, $[z]$, $[z,z\!+\!1]$), or \emph{orthogonal reslices} (XY/XZ/YZ). Replace dims as appropriate.}
\label{fig:architecture}
\end{figure*}
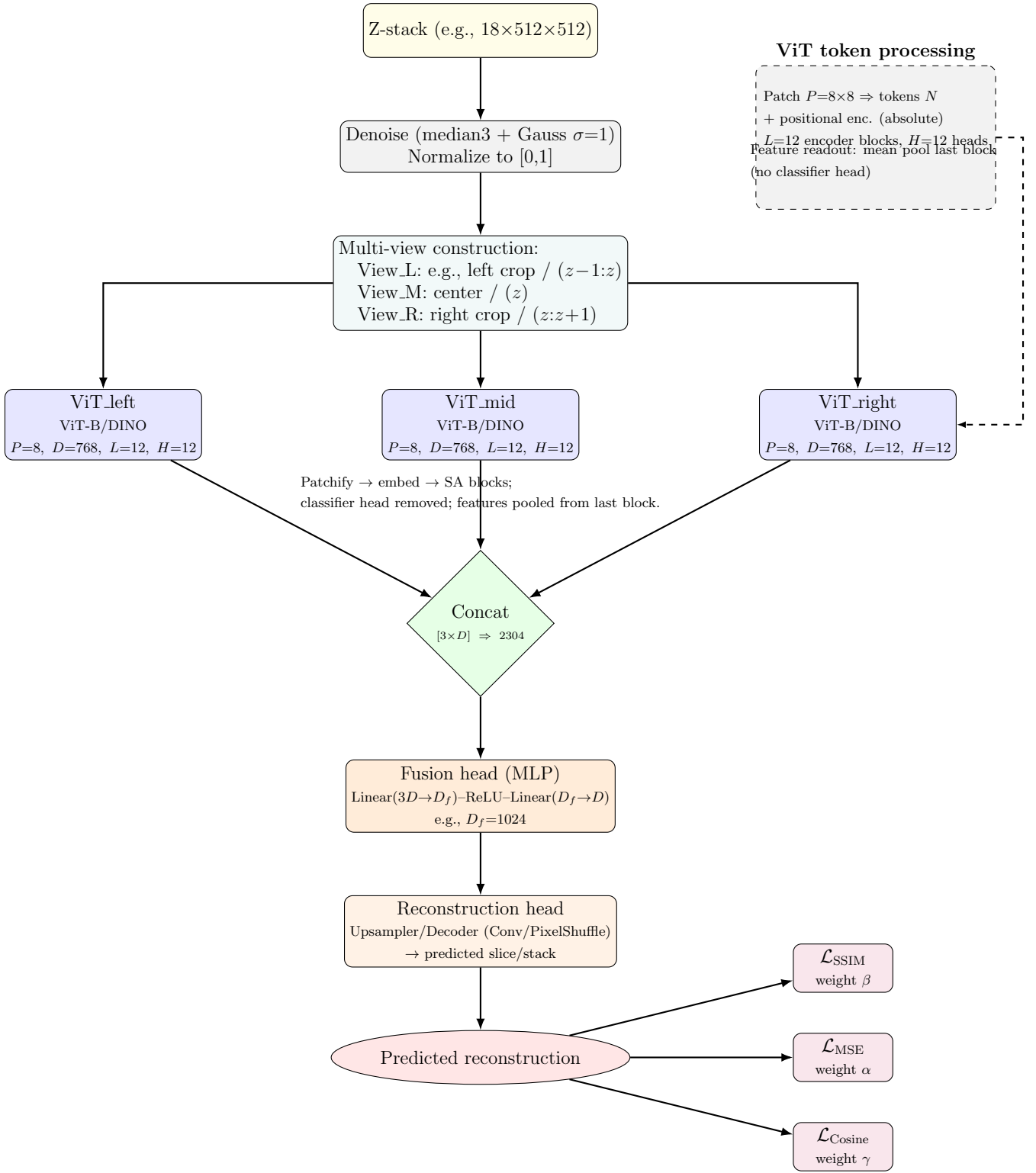

We use Vision Transformers (ViTs) as the core of our neural network because they have been shown to be effective at processing spatial and timing patterns \cite{dosovitskiy2020image}. Unlike convolutional neural networks that use local receptive fields, ViTs use self-attention to model long-range dependencies in visual data \cite{raghu2021vision}. This global processing lets ViTs capture complex spatial relationships and timing patterns critical for 3D+T data. Specifically, we use "vit\_base\_patch8\_224\_dino" \cite{rw2019timm} model pre-trained using the DINO method, priming them with robust visual representations, shown visually in Figure \ref{fig:architecture}. DINO's focus on aligning features during knowledge transfer readies these models for our space and time prediction challenges \cite{caron2021emerging}. Overall, ViTs provide a strong backbone network matched to the complexity of our cell culture image data.

Our design uses three distinct ViTs - vit\_mid, vit\_left, and vit\_right for feature extraction. From each Z-slice we extract three overlapping lateral crops (left/middle/right; 70\% field-of-view, 20\% overlap). Each crop feeds one ViT branch; features are fused channel-wise. Each ViT takes in the input 3D+T images and extracts space-time representations without the classification portion. This utilizes the full feature richness encoded by the ViTs, not just the final class predictions. Using three ViTs to process the data from different views provides wider coverage of the varied patterns and clues within the detailed cell culture images. Their outputs are combined through later fusion layers, enabling an integrated feature representation key for reliable 3D+T forecasting.

The feature representations extracted by vit\_mid, vit\_left, and vit\_right offer complementary views into the complex attributes of the 3D+T data. To combine these varied space-time clues into a unified representation, we use fusion layers to assimilate the individual ViT outputs. Specifically, the fusion layers integrate learned linear transformations coupled with ReLU (Rectified Linear Unit) activations to merge and extract the key aspects from each ViT into a consolidated feature set. This integration promotes a comprehensive representation key for reliable 3D+T forecasting, bringing together the strengths of each ViT's specialized perspective.

Our fusion layers perform weighted aggregations of the feature maps extracted by vit\_mid, vit\_left, and vit\_right. Strategically combining these complementary representations is vital for forming a complete depiction of the complex spatial details and timing patterns. The fusion uses linear transformations followed by ReLU activations. The linear transformations learn optimal weighting tailored to the importance of each ViT's feature set for predicting the target 3D+T frames. Meanwhile, the non-linear ReLU units enrich the expressiveness of the combined features. Together, the thoughtful fusion integrates the varied space-time clues into a unified comprehensive representation. Our work is available on \textit{GitHub}, implemented in our public codebase~\cite{Ugurlar2025VTDTXN}.

\subsubsection{Custom Loss Function Formulation}

We formulate a tailored loss function to guide the model's training process for our unique 3D+T prediction challenges:

\begin{equation}
\mathcal{L}(\mathbf{y}, \hat{\mathbf{y}}) = \alpha \mathcal{L}{\mathrm{MSE}}(\mathbf{y}, \hat{\mathbf{y}}) + \beta \mathcal{L}{\mathrm{SSIM}}(\mathbf{y}, \hat{\mathbf{y}}) + \gamma \mathcal{L}_{\mathrm{Cosine}}(\mathbf{y}, \hat{\mathbf{y}})
\label{eq:loss}
\end{equation}

This loss combines MSE to measure pixel-level accuracy, SSIM to evaluate structural similarity vital for 3D+T images, and Cosine Similarity to assess feature alignment. The coefficients \(\alpha\), \(\beta\), and \(\gamma\) are tuned to balance these terms' contributions in line with our desired training goals. This customized formulation maintains agreement between our loss function and evaluation metrics, steering the model toward predictions that are not just numerically accurate but also visually and structurally meaningful.

\subsubsection{Model Optimization}
To optimize for efficient inferencing, we use pruning and quantization strategies: (i) Pruning removes redundant or unimportant connections, reducing model size. We use magnitude-based pruning, removing low-weight connections first. (ii) Quantization lowers the precision of weights and activations. We apply 8-bit quantization with minimal accuracy loss.Together these methods extract a compact yet accurate model. Pruning reduces parameters and quantization shrinks memory use to streamlined models specialized for our prediction tasks \cite{kim2023quantization, kim2021pqk}.

Optimizing for inference enables researchers to analyze cell cultures during experiments. We accelerate inference through:

\begin{itemize}
    \item Model pruning to minimize computational operations.
    \item Weight quantization to enable faster 8-bit math operations.
    \item Batch optimizations like fusion to speed up batch processing.
    \item Hardware acceleration using GPUs, FPGAs, or dedicated ASICs.
    \item Streamlining software libraries like onnx for lean inference.
\end{itemize}
By profiling speed and targeting bottlenecks, we can tune the model for near real-time turnaround. This unlocks real-time cell analysis to guide interventions and decisions even during experiments \cite{li2022distributed, sharify2019laconic}.

\subsubsection{Training and Optimization}

To stabilize training, we use gradient accumulation to mimic larger batch sizes. This lessens gradient noise and produces more steady model updates \cite{lamy2021layered, andersson2022end}. We also apply a ReduceLROnPlateau scheduler to dynamically adjust the learning rate based on the validation loss plateauing \cite{al2022scheduling}. This fine-tunes the training pace, preventing divergent oscillations or stalling during optimization. Together, these strategies smooth and speed up training convergence for our high-dimensional 3D+T data.

We use regularization techniques like early stopping and dropout to prevent overfitting. Early stopping halts training when validation metrics plateau, avoiding over-specializing on the training data. Meanwhile, dropout randomly omits units during training, making the model robust to missing inputs. This combination provides a system of checks and balances, enabling the model to reliably generalize to unseen data \cite{finnoff1993improving, JMLR:v15:srivastava14a}.

We chose the Adam optimizer to guide model training due to its adaptive learning rate and momentum mechanisms. By independently tuning the learning rate for each parameter based on magnitude and variance estimates, Adam speeds up convergence consistency \cite{kingma2014adam}. Additionally, its momentum integration helps coast over small local optima. Together, these properties make Adam well-suited for the high-dimensional optimization landscape of our 3D+T prediction problem.

\subsubsection{Evaluation Methodology}

To thoroughly evaluate our model, we chose metrics mirroring our custom loss function, ensuring alignment between training and evaluation. Each metric provides a unique perspective on the predictions:

\begin{itemize}
    \item \textbf{Mean Squared Error (MSE):} A fundamental metric in regression analysis, the MSE computes the average squared differences between the model's predictions and the ground truth \cite{schluchter2005mean, furnkranz2010mean}. Mathematically, it is represented as:
    
    \begin{equation}
    \text{MSE} = \frac{1}{n} \sum_{i=1}^{n} (y_{i} - \hat{y}_{i})^2
    \label{eq:MSE}
    \end{equation}

    \item \textbf{Structural Similarity Index (SSIM):} Beyond mere pixel-level accuracy, the visual structure and patterns in the predictions matter, especially in the context of 3D + T cell culture imagery \cite{nilsson2020understanding, wang2004image, bakurov2022structural, renieblas2017structural}. The SSIM is computed as:
    
    \begin{equation}
    \text{SSIM}(x, y) = \frac{(2 \mu_x \mu_y + c_1)(2 \sigma_{xy} + c_2)}{(\mu_x^2 + \mu_y^2 + c_1)(\sigma_x^2 + \sigma_y^2 + c_2)}
    \label{eq:SSIM}
    \end{equation}
    
    \item \textbf{Cosine Similarity (CS):} A geometric perspective on similarity, this metric assesses the cosine of the angle between two non-zero vectors \cite{xia2015learning}. Mathematically, it's given by:
    
    \begin{equation}
    \text{cosine similarity} = \frac{\mathbf{A} \cdot \mathbf{B}}{||\mathbf{A}||_2 \times ||\mathbf{B}||_2}
    \label{eq:CosineSimilarity}
    \end{equation}
\end{itemize}

\section{RESULTS}

We provide a range of results based on a comprehensive performance analysis that provides a balance overview of how the VT-DTSM in terms of error analysis (MSE metrics), pixel level quality (SSIM) as well as visual features of the stacked images as vectors (CS).

\begin{figure*}[t]
\centering

\begin{minipage}[t]{0.34\textwidth}\vspace{0pt}
  \centering
  \includegraphics[width=\linewidth]{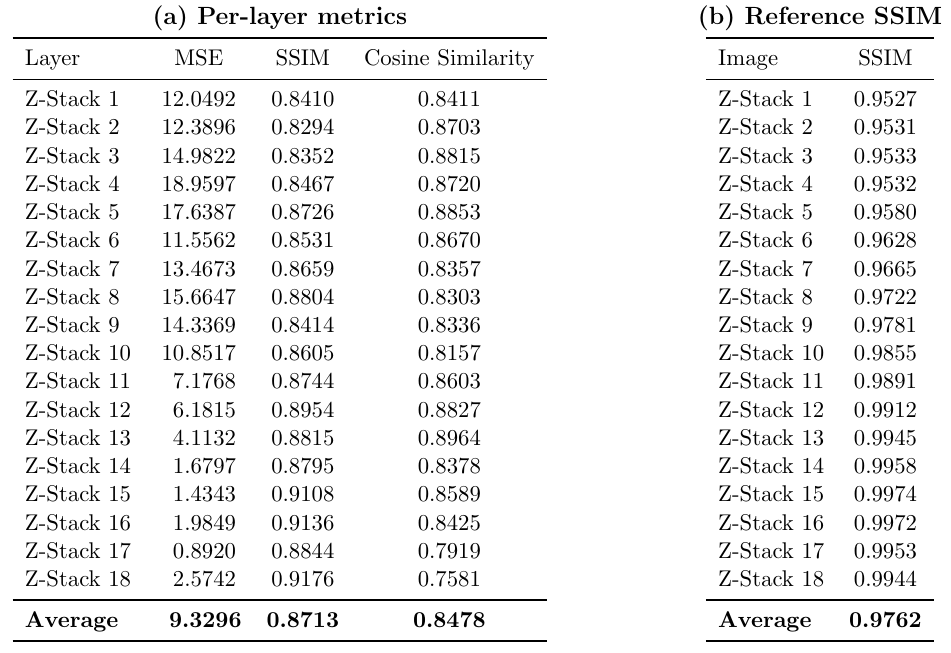}

  \vspace{0.6em} 

  \includegraphics[width=\linewidth]{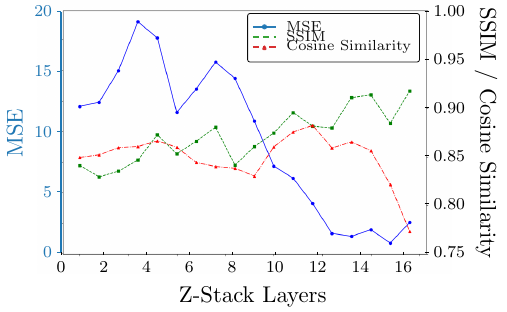}
\end{minipage}
\hfill
\begin{minipage}[t]{0.34\textwidth}\vspace{0pt}
  \centering
  \includegraphics[width=\linewidth]{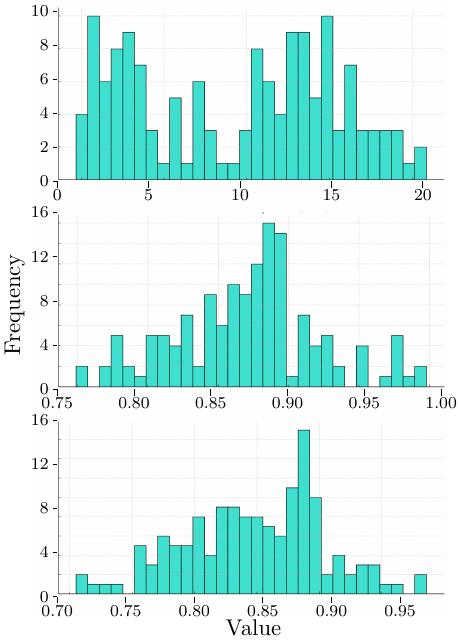}
\end{minipage}
\hfill
\begin{minipage}[t]{0.28\textwidth}\vspace{0pt}
  \centering
  \includegraphics[width=\linewidth]{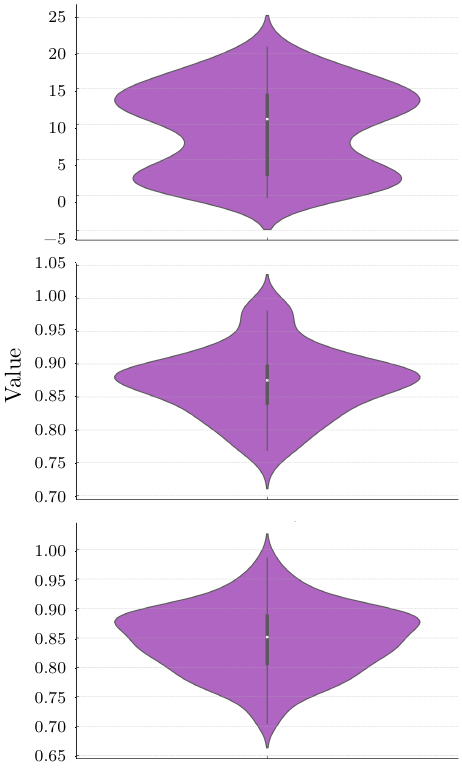}
\end{minipage}

\caption{Comprehensive quantitative evaluation of the Vision Transformer Digital Twin Surrogate Network (VT-DTSN) for 3D+T \textit{Drosophila} midgut reconstruction across 18 Z-stack layers and multiple samples. (Top-left) Layer-wise MSE, SSIM, and Cosine Similarity metrics. (Middle-right) Histograms displaying the distribution of MSE, SSIM, and Cosine Similarity. (Far-right) Violin plots showing the spread and consistency of MSE and SSIM across the dataset. (Bottom-left) Line plots tracking MSE (blue), SSIM (green), and Cosine Similarity (red) across Z-stack layers, demonstrating consistent structural preservation and feature alignment despite depth-dependent variability.}
\label{fig:results_all}
\end{figure*}

\begin{figure}
\centering

\begin{tikzpicture}
  \node[anchor=south west, inner sep=0] (image)
    at (0,0) {\includegraphics[width=70mm]{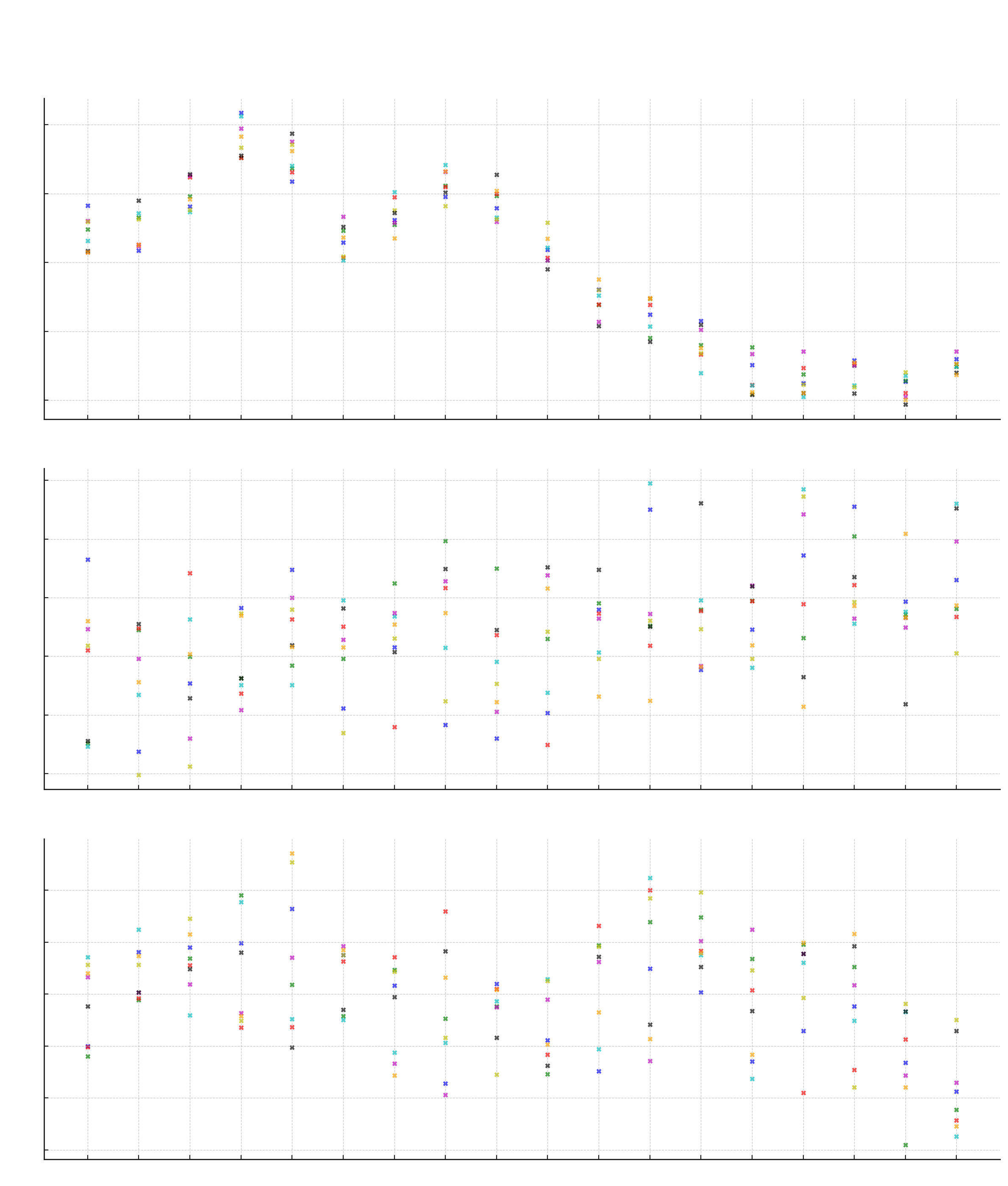}};

  \begin{scope}[x={(image.south east)}, y={(image.north west)}]

    \newcommand{\tickfont}{\fontsize{8.5}{10}\selectfont}
    \newcommand{\axisfont}{\fontsize{10.5}{12}\selectfont}
    \def\ticklen{0.010}
    \def\xtickyoff{0.008}   
    \def\ytickxoff{0.004}   

    \def\xL{0.048} \def\xR{0.947}
    \def\yTOne{0.915} \def\yBOne{0.650}  
    \def\yTTwo{0.590} \def\yBTwo{0.336}  
    \def\yTThr{0.275} \def\yBThr{0.021}  


    \newcommand{\xtick}[6]{\pgfmathsetmacro\x{#4+(#1-#2)*(#5-#4)/(#3-#2)}%
      \draw (\x,#6) -- (\x,#6-\ticklen);}
    \newcommand{\xticklabel}[6]{\pgfmathsetmacro\x{#3+(#1-#2)*(#4-#3)/(#5-#2)}%
      \node[anchor=north] at (\x,#6-\xtickyoff) {\tickfont \num{#1}};}
    \newcommand{\ytick}[6]{\pgfmathsetmacro\y{#4+(#1-#2)*(#5-#4)/(#3-#2)}%
      \draw (#6-\ticklen,\y) -- (#6,\y);}
    \newcommand{\yticklabel}[6]{\pgfmathsetmacro\y{#3+(#1-#2)*(#4-#3)/(#5-#2)}%
      \node[anchor=east] at (#6-\ytickxoff,\y) {\tickfont \num{#1}};}

    \def\YTicks{0,5,10,15,20}                       
    \def\XTicks{0,1,2,3,4,5,6,7,8,9,10,11,12,13,14,15,16,17}              
    \def\YTicksSSIM{0.75,0.80,0.85,0.90,0.95,1.00}  
    \def\YTicksCos{0.70,0.75,0.80,0.85,0.90,0.95}   

    \def\xmin{0}\def\xmax{17}  

    \def\yminA{0}\def\ymaxA{21}
    \foreach \t in \XTicks { \xtick{\t}{\xmin}{\xmax}{\xL}{\xR}{\yBOne} }
    \foreach \t in \YTicks  {
      \ytick{\t}{\yminA}{\ymaxA}{\yBOne}{\yTOne}{\xL}
      \yticklabel{\t}{\yminA}{\yBOne}{\yTOne}{\ymaxA}{\xL}
    }

    \def\yminB{0.75}\def\ymaxB{1.00}
    \foreach \t in \XTicks { \xtick{\t}{\xmin}{\xmax}{\xL}{\xR}{\yBTwo} }
    \foreach \t in \YTicksSSIM {
      \ytick{\t}{\yminB}{\ymaxB}{\yBTwo}{\yTTwo}{\xL}
      \yticklabel{\t}{\yminB}{\yBTwo}{\yTTwo}{\ymaxB}{\xL}
    }

    \def\yminC{0.70}\def\ymaxC{0.98}
    \foreach \t in \XTicks {
      \xtick{\t}{\xmin}{\xmax}{\xL}{\xR}{\yBThr}
      \xticklabel{\t}{\xmin}{\xL}{\xR}{\xmax}{\yBThr}
    }
    \foreach \t in \YTicksCos {
      \ytick{\t}{\yminC}{\ymaxC}{\yBThr}{\yTThr}{\xL}
      \yticklabel{\t}{\yminC}{\yBThr}{\yTThr}{\ymaxC}{\xL}
    }

    \node[anchor=north] at (0.53,-0.028) {\axisfont Sample Index};
    \node[anchor=south, rotate=90] at (-0.052,0.50) {\axisfont Frequency / Value};

  \end{scope}
  
\end{tikzpicture}

    \caption{Scatter plots depicting the average values of the eight distinct sample sets for the Mean Squared Error (MSE), Structural Similarity Index (SSIM), and Cosine Similarity.}
    \label{fig:scatter_plot}
\end{figure}
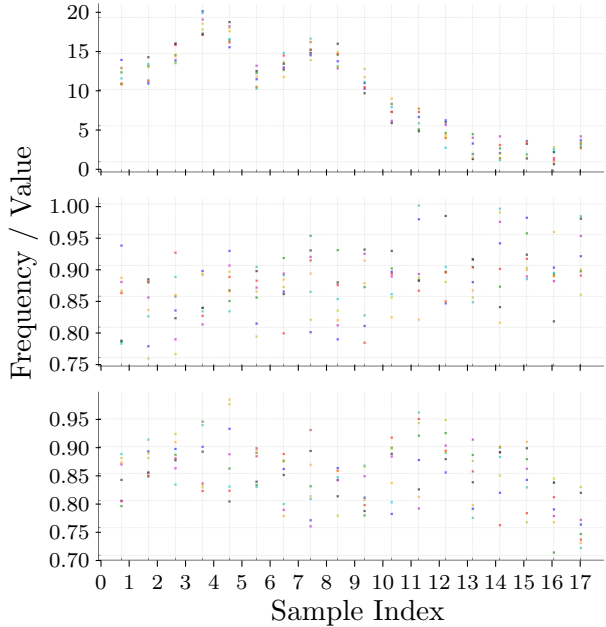

\subsection{Pixel-Level Reconstruction Fidelity Across Depth}

A critical requirement for reliable 3D+T SDTMs is the preservation of pixel-level detail across the Z-stack, enabling accurate morphological quantification. We evaluated pixel fidelity using the Mean Squared Error (MSE) across all 18 Z-stack layers and eight biological replicates, as presented in Figure \ref{fig:results_all} and Figure \ref{fig:scatter_plot}. Figure \ref{fig:results_all} includes a layer-wise table of MSE values, histograms indicating the distribution of errors across the dataset, a violin plot showing variance, and a depth-wise performance plot. These visualizations demonstrate that MSE increases moderately in deeper layers, with values ranging from $0.89$ in $Z17$ to $18.95$ in $Z4$, reflecting the increased complexity of imaging in deeper tissue regions while maintaining overall stable error profiles across replicates. Figure \ref{fig:scatter_plot} expands on this analysis, providing scatter plots across layers and samples, demonstrating consistent trends and confirming the absence of significant outlier behaviours across biological replicates. This analysis indicates that the model maintains reliable pixel-wise accuracy throughout the Z-stack, enabling the faithful capture of cellular structures critical for downstream segmentation, measurement, and quantitative morphological analyses in dynamic surrogate model systems.

\subsection{Structural and Perceptual Integrity}

In addition to pixel-level fidelity, structural and perceptual consistency across layers is essential to preserve biologically meaningful features within reconstructed images. We assessed structural fidelity using SSIM, which captures luminance, contrast, and structural consistency. As shown in Figure \ref{fig:results_all}, SSIM values remain high across all Z-stack layers, with averages ranging from $0.84$ to $0.91$, demonstrating low variance across samples and layers. The histogram displays a distribution skewed toward high similarity, while the violin plot confirms stability across replicates. The depth-wise trend in the performance plot shows that SSIM remains stable, with minor improvements in deeper layers where low-frequency structures dominate. Figure \ref{fig:scatter_plot} confirms this pattern across biological replicates, showing high SSIM values across all layers and samples. These results confirm that the VT-DTSM preserves perceptual and structural integrity, supporting interpretability for biological experts and ensuring the reconstructed digital twins as a surrogate model can be reliably used for assessing tissue architecture and cellular morphology across time and depth.

\subsection{Feature-Space Alignment Across Layers}

Maintaining consistency in feature representations across the Z-stack is crucial for downstream tasks such as cell-type classification, segmentation, and feature-based tracking. We evaluated feature-space alignment using CS between predicted and ground truth representations. In Figure \ref{fig:results_all}, CS values are consistently high across all layers, averaging between $0.84$ and $0.87$. The histogram and violin plot illustrate a narrow, high-similarity distribution, and the performance plot shows only minor decreases in deeper layers where complexity and noise increase. Figure \ref{fig:scatter_plot} complements this analysis, providing scatter plots demonstrating consistent feature alignment across all biological replicates and layers. The VT-DTSM model retains meaningful high-level representations across depth, ensuring the digital twins can be integrated seamlessly into advanced machine learning pipelines for further analysis without losing critical biological information.

\subsection{Qualitative Analysis}

\begin{figure*}
\floatplacement{figure}{H}
\centering
\includegraphics[width=0.74\textwidth]{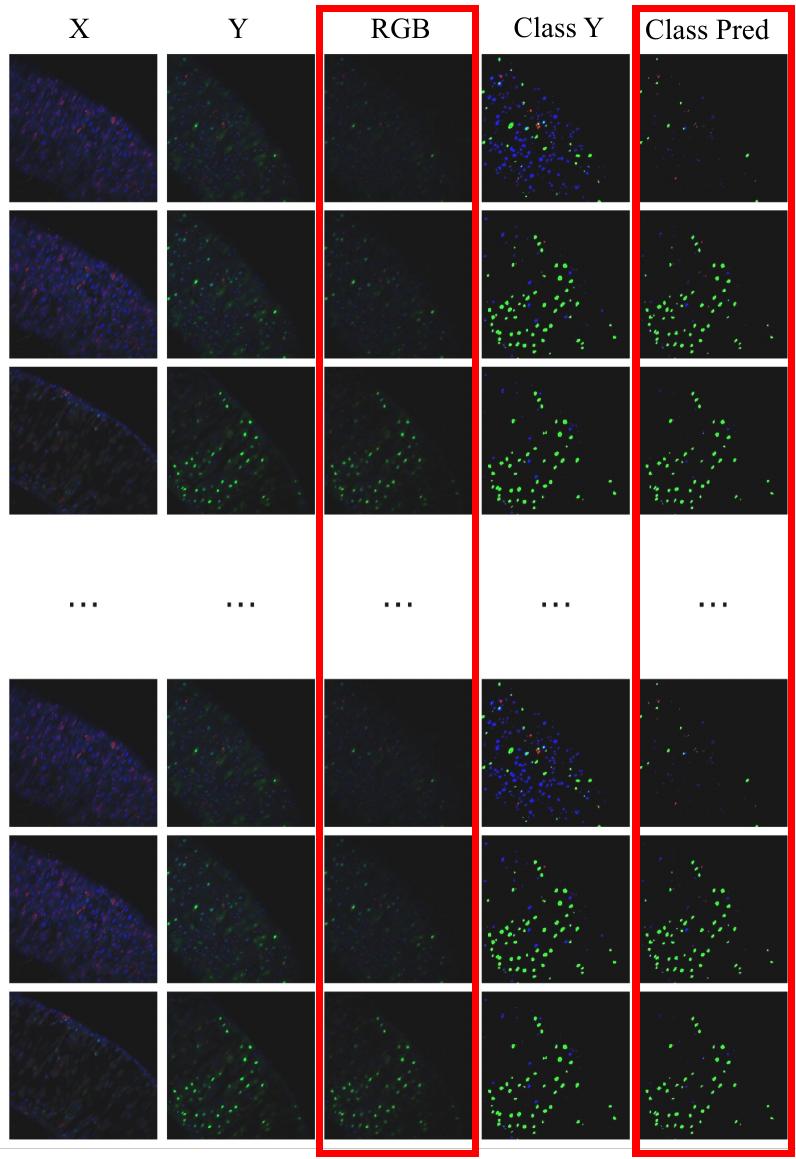}
\caption{Comparison of the ground truth and predicted images of the sample 1. From left to right: Ground Truth X Value, Ground Truth Y Value, Predicted RGB Picture, Ground Truth 4 Class Picture, Predicted 4 Class Picture.}
\label{fig:image_grid}
\end{figure*}

Figure \ref{fig:image_grid} presents representative frames illustrating the input images, ground truth labels, and the corresponding predicted outputs generated by the Vision Transformer-based model across three representative samples. The first two columns in Figure \ref{fig:image_grid} display the raw fluorescence input channels (X) and ground truth labels (Y), showing the spatial distribution of cellular structures across the midgut tissue. The third column presents the reconstructed RGB predictions, demonstrating the model’s capacity to recover fine morphological details, including cellular boundaries, nuclear regions, and the layered organization of the midgut epithelium. Visually, the predicted outputs align closely with the ground truth across all samples, with no evident blurring or spatial artifacts, even in regions of high cellular density.

The last two columns in Figure \ref{fig:image_grid} show the class-specific overlays for the ground truth (Class Y) and predicted labels (Class Pred). Here, differentiated cell types and marker-based classes are distinctly preserved, with clear spatial separation and accurate localization of individual cells. The predicted class distributions closely mirror the ground truth, with the model capturing the correct positioning and density of marker-positive cells across the tissue, demonstrating its ability to retain class-specific biological features during reconstruction.

This qualitative validation confirms that the VT-DTSM can generate high-fidelity reconstructions that accurately replicate the spatial complexity of the midgut tissue. Reconstructions preserve epithelial boundaries and lumen contours across depth; we did not observe spurious merging or tearing artifacts in high-density regions on the held-out replicates. The ability to preserve morphological features and class-specific cellular distributions is critical for downstream biological interpretation, enabling accurate assessments of proliferation, cell type distribution, and tissue organization in live-imaging or computational experiments. Importantly, the visually interpretable results support the model’s potential for integration into experimental workflows where rapid and reliable assessments of midgut dynamics are required, without compromising the biological relevance of the data.

\section{DISCUSSION}

The VT-DTSN demonstrates the feasibility of using Vision Transformers as high-fidelity digital twin surrogates for the \textit{Drosophila} midgut, enabling accurate reconstruction of spatial and temporal tissue dynamics from 3D+T imaging data. Compared to conventional CNN-based approaches, our method captures long-range spatial dependencies critical for understanding layered epithelial structures, while our use of DINO-pretrained models with a multi-view fusion strategy ensures robust feature learning across imaging depths. The resulting model not only reconstructs morphological features with low error and high structural similarity but also retains feature-level consistency essential for downstream analyses such as cell tracking, segmentation, and lineage inference. By enabling in silico experimentation, the VT-DTSN reduces experimental load, accelerates hypothesis testing, and allows researchers to explore perturbation effects in a virtual environment before live validation, which is a significant advance over current static analysis pipelines in midgut research.

Despite these advantages, our approach has some considerations. The VT-DTSN’s performance can be influenced by domain-specific imaging conditions, such as variable signal-to-noise ratios or imaging artifacts at greater depths, which may affect generalization to datasets acquired under different conditions or with alternative imaging modalities. In this dataset, timepoints are cross-sectional across independent specimens; ‘temporal’ therefore denotes variability across replicates rather than longitudinal trajectories of the same midgut. While our model preserves morphological and structural fidelity, it currently does not incorporate explicit biological constraints such as cell lineage or signaling dynamics, which would enhance its interpretability and predictive power. VT-DTSN is a data-driven surrogate and does not explicitly encode mechanistic priors (e.g., lineage, signaling, biomechanics). Future work will focus on integrating multi-channel marker data, domain adaptation techniques for cross-lab generalization, and coupling the VT-DTSN with agent-based or graph-based models to capture cell-cell interactions and mechanical constraints. These integrations will move the digital twin from a high-fidelity reconstructive surrogate toward an interactive, mechanistically grounded model that complements live imaging to advance the understanding of epithelial biology in \textit{Drosophila} and beyond.

\section{CONCLUSION}
In this study, we have developed and validated the Vision Transformer Digital Twin Surrogate Network (VT-DTSN), demonstrating its ability to generate accurate, high-fidelity predictive reconstructions of dynamic 3D+T imaging data from extracted \textit{Drosophila} midgut tissue, providing a feasible, high-fidelity surrogate for cross-timepoint reconstruction, for investigating cellular dynamics and tissue homeostasis in silico while complementing live imaging workflows.

\bibliographystyle{IEEEtran}
\bibliography{ref}


\end{document}